\documentclass{elsarticle}
\usepackage{lineno,hyperref}
\usepackage{amsfonts}
\modulolinenumbers[5]

\journal{Physics Letters B}

\begin{document}

\begin{frontmatter}

\title{Actions, equations of motion and boundary conditions for Chern-Simons Branes}

\author{Pablo Mora}
\address{Centro Universitario Regional Este (CURE), Universidad de la Rep\'ublica, Uruguay, Ruta 9 km. 207, Rocha, Uruguay}

\ead[url]{www.cure.edu.uy}

\corref{mycorrespondingauthor}
\cortext[mycorrespondingauthor]{Corresponding author}
\ead{pablomora@cure.edu.uy}

\begin{abstract}
Actions for extended objects based on Transgression and Chern-Simons forms for space-time groups and supergroups provide a gauge 
theoretic framework in which to embed previously studied String and Brane actions, 
extending them in interesting ways that may be useful in a future non perturbative formulation of String Theory. 
In this Letter I investigate aspects of the 
actions of these theories, including equations of motion and boundary conditions, gauge and space-time symmetries, and Dirac quantization of tensions. 
This theoretical framework is shown to include in certain limit and for a suitable gauge group the standard Bosonic String Theory.  
\end{abstract}


\end{frontmatter}

\section{Introduction}

Actions for extended objects (branes) based on Chern-Simons and Transgression forms for space-time groups were introduced in \cite{Mora-Nishino20001,Mora20011} by H. Nishino and myself, aiming to build a framework in which previous string and brane models could be embedded, endowing them with desirable additional features. Those models are background independent gauge systems for the space-time groups. The background is not fixed and given a priori, but participates of the dynamics of the extended objects, which are sources of the gauge fields. 
The actions of the branes have the same mathematical form than that of the background, in what we called "brane-background democracy". 

The above mentioned work built on, and was inspired by, several previous lines of research:
\begin{enumerate}[(I)]
\item It introduces extended objects in a gauge invariance way in Chern-Simons (super)gravity theories \cite{Achucarro:1987vz,Chamseddine:1990gk, Chamseddine:1989nu, MuellerHoissen:1990vf, Banados:1993ur, Banados:1996hi, Zanelli:2012px}, which had 
the goal of extending General Relativity to describe gravity as a gauge theory for some space-time (super)group. 
\item It generalizes Dixon et al. (DDS) models \citep{Dixon:1991xz} for the coupling of 
extended objects to non abelian gauge fields (heterotic branes) to space-time groups and their supersymmetric extensions (treated in the same footing as other possible gauge groups). It differs from the DDS models in that the background described is described by Chern-Simons/Transgression gravity or supergravity, in a way formally analogous to the extended objects themselves, and not by standard gravity or supergravity; and in the fact that our models allow both sets of gauge fields involved to be non pure gauge. 
\item It was motivated by Green's construction \cite{Green:1989nn} of the
superstring action as a gauged Wess-Zumino-Witten (WZW) model for a supergroup.
\item It was also motivated by the Chern-Simons supergravity theory for the supergroup $OSp(32,1)$ proposed by Troncoso and Zanelli \cite{Troncoso1997va, Troncoso1998ng} and its conjectured relationship with eleven dimensional supergravity\cite{Cremmer:1978km}, and by Horava's work \cite{Horava:1997dd} suggesting a possible relation between a Chern-Simons supergravity with gauge group $OSp(32,1)\times OSP(32,1)$ and M-theory \cite{Hull:1994ys,Witten:1995ex,Duff:1995wd}.

\end{enumerate}

In the following years there were several works exploring diverse aspects of models of Chern-Simons branes \cite{Edelstein:2008ry,Miskovic:2009dd,Edelstein:2010sx,Edelstein:2010sh,Edelstein:2011vu,Kastikainen:2020auf,Frey:2019fqz,Ertem:2012qv}, including explicit solutions and conserved charges.

In Ref.\cite{Mora-Nishino20001} we suggested that a model of this class may be useful for of a non perturbative description of String Theory, or M-theory. I still believe that possibility is the main reason to study these models. Some of the motivations to do so, and the features that distinguish them from other extended objects models are:
\begin{enumerate}[(i)]

\item As stated in Ref.\cite{Mora-Nishino20001}\footnote{The quote is \cite{Mora-Nishino20001}: "A heuristic motivation for the kind of models presented here is that if there is some
underlying pregeometric theory in which the de Rham complex get physical content through
some kind of antighost field variables, we must expect the effective action to be a sum
of pieces of diverse dimensions of the form considered here. We believe that this simple
observation is the key to understand the existence of extended objects in a fundamental
theory of nature."}, considering an action that is a sum of integrals of differential forms of different orders but similar mathematical form \footnote{One may think of the action as a sort of generalized integral of a polynomial, or truncated expansion (as befits to Grassmann variables), on the differentials $dx^{\mu}$.} can be understood as a way to give physical content to the differential structure of the manifold, as General Relativity had given physical content to the metric structure,  giving another way to justify the possible existence of fundamental extended objects. The frequent question "what are strings made of?" is often answered with the question "what are particles made of?", somehow implying the first question is moot, yet from our point of view the answer may be a key to the foundations of String Theory.

\item The only dynamical fields are gauge fields, although some of them are fermionic in the case 
of a gauge supergroup, providing the fermions required for a connection with Superstrings. 
The p-form fields that would be associated to those occurring in String Theories are composite, 
and given by Chern-Simons forms of our gauge fields, what means that the anomalous 
gauge transformations and Bianchi identities of those fields are built-in. 
The notion that the fundamental degrees of freedom of String Theory may not be the apparent ones has been widely accepted for quite some time, see for instance Ref.\cite{Witten:1988sy}\footnote{Quoting from this Reference: "In my opinion, the basic challenge in string theory is not, as sometimes said, to 'understand
non-perturbative processes in string theory'. In fact, the basic problem does not lie in the
quantum domain at all. The basic problem, now and probably for many years to come, is to
understand the classical theory properly."  And "So the central task is to
(i) find the right degrees of freedom;
(ii) find the right invariance group; and
(iii) formulate the right lagrangian."}.
 In the field theory case it has been found that full gauge invariance (instead of quasi-invariance) leads from Chern-Simons forms to Transgressions, which automatically dictates the correct boundary terms to regularize Noether conserved charges and black hole thermodynamics \cite{MOTZ-2004,MOTZ-2006,Mora20141}. This can be taken as evidence for requiring a gauge invariant background independent theory being the right path to follow.
 
\item These theories are already first quantized, analogous to strings on non-flat backgrounds. They differ from those string models in that the background is described by  Chern-Simons  (super)gravity instead of standard (super)gravity. 
It is remarkable that the above mentioned theory with supergroup $OSp(32,1)$ of Troncoso and Zanelli \cite{Troncoso1997va, Troncoso1998ng} has been shown by Izaurieta and Rodr\'{\i}guez \cite{Izaurieta:2011fr} to reduce in certain limit to a theory very similar to standard eleven dimensional supergravity \cite{Cremmer:1978km}, with corrections given by a expansion in powers of the parameter used in the limit. It seems possible that with further analysis of the correspondences between the fields of both theories, and a suitable choice of the invariant trace involved, the correspondence may be more precise, and the additional terms could be coherent with anomalous M-theoretic corrections from dualities \cite{Duff:1995wd}.

\item In a sort of holographic property, if the gauge fields are pure gauge then only degrees of freedom at the boundaries of the branes and the space-time remain.

\end{enumerate}

The plan of this Letter is the following:
I review the issue of the equations of motion and boundary conditions, both in the case of generic and pure gauge connections, and the symmetries of these theories. 
I discuss some additional conditions, leading to desirable properties on these models. These include the inclusion of the kinetic term, added at the boundary and its coefficient relative to the bulk Chern-Simons term in the bosonic case, independently of the argument based in kappa-symmetry given in \cite{Mora-Nishino20001}, as well as the exclusion of additional possible bulk terms.
Afterwords I consider the particular case or sub-sector of the theory that makes contact with the Bosonic String in some limit.
Finally I discuss the existence of Dirac quantization conditions on the tensions or coupling constants of the branes.

\section{Transgression Actions for Branes}

I consider actions for extended objects (branes) with dynamical variables given by gauge fields given as differential 1-forms $A=A_{\mu}^AG_Adx^{\mu}$ valued on the gauge group $\mathcal{G}$ algebra with generators $G_A$, with $\mu =0,...,D$, with $D$ even, the embedding coordinates of each brane 
$X_{(d+1)}^{\mu}(\xi _{(d+1)}^i)$, where $\xi _{(d+1)}^i$, 
with $i=0,...,d$, with $d$ even, are the coordinates of the $d$ brane, and the intrinsic metrics defined at the boundaries of the branes $\gamma _{(d)rs}$, with $r,s=0,...,d-1$. The space-time in which 
the branes are embedded can be consider itself as a brane, and is itself described by an action of 
the same class as the actions of the branes, therefore 
its coordinates are $\xi _{(D+1)}^{\mu}\equiv x^{\mu}$). While the branes themselves may or may not have boundaries, the boundaries of the branes have no boundary. 
The actions considered, introduced in Ref.\citep{Mora-Nishino20001}, are given as a sum of terms of diverse dimensions corresponding to the various branes and the space-time in which these branes are embedded 
\begin{equation}
S=\sum _{d=2, ~d~even}^{D}[S_{d+1}^{(Trans)}+S_{d}^{(Kin)} ],
\end{equation}
where the Transgression action part is given by
\begin{equation}
S_{d+1}^{(Trans)}=\alpha _{d+1}\int _{\mathcal{S}^{d+1}} \mathfrak{T}_{d+1},
\end{equation}
with the Transgression forms given by 
\begin{equation}
\mathfrak{T}_{d+1}(A,\overline{A})=[\frac{d}{2}+1]\int_0^1dt<\Delta AF_t^{\frac{d}{2}}>.
\end{equation}
There $\Delta A=A-\overline{A}$, $A_t=tA+(1-t)\overline{A}$ and $F_t=dA_t+A_t^2$. The brackets $<...>$ denote symmetrized (super)traces on the algebra of the gauge (super)group, corresponding to symmetric invariant tensors $g_{A_1...A_{d}}\equiv <G_{A_1}...G_{A_{d}}>$. These symmetrized traces or invariant tensors, for which the simplest choice is $<...>=STr[...]$, are not unique. Through all this work we use a notation where the exterior (wedge) product of differential forms is implied 
$\alpha _p\wedge \beta _q\equiv \alpha _p\beta_q$.

The Kinetic parts of the action has support on the boundaries of the 
branes, if such boundaries exist, and are of the form
\begin{equation}
S_{d}^{(Kin)}=\beta _d\int _{\mathcal{S}^d\equiv \partial \mathcal{S}^{d+1}}d^d\xi _{(d)}
\sqrt{-\gamma _{(d)}} ~\gamma _{(d)}^{ij}[<\Delta A_i\Delta A_j>-(d-2)]
\end{equation}
where $\Delta A_i=\Delta A_{\mu}^A\frac{\partial X_{(d)}^{\mu}}{\partial \xi _{(d)}^i}G_A$.

\section{Equations of motion}

\subsection{Equations of motion associated to the variations of the gauge fields}

Under variations of $A$ and $\overline{A}$ the variation of the Transgression part of the action is
\begin{eqnarray}
\delta _{A,\overline{A}} S_{d+1}^{(Trans)}=\sum_{d=2,~ d~even}^D\alpha _{d}
\int _{\mathcal{S}^{d+1}}(\frac{d}{2}+1)
[ <F^{\frac{d}{2}}\delta A>-<\overline{F}^{\frac{d}{2}}\delta \overline{A}> ]-\nonumber\\
-\sum_{d=2,~d~even}^N\alpha _d\int _{\mathcal{S}^{d}\equiv \partial \mathcal{S}^{d+1}}
\frac{d}{2}(\frac{d}{2}+1)\int _0^1 dt 
<\Delta A F_t^{\frac{d}{2}-1}\delta _{A,\overline{A}} A_t>,
\end{eqnarray}
where $\delta _{A,\overline{A}} A_t=t\delta A +(1-t)\delta \overline{A}$. 

The variations of the kinetic terms are
\begin{equation}
\delta _{A,\overline{A}} S_{d}^{(Kin)}=2\beta _d\int _{\mathcal{S}^d\equiv \partial \mathcal{S}^{d+1}}d^d\xi _{(d)}
\sqrt{-\gamma _{(d)}} ~\gamma _{(d)}^{ij}<\Delta A_i(\delta A_j  - \delta \overline{A}_j  )>.
\end{equation}
We will need that using Dirac deltas we can rewrite
\begin{eqnarray}
\int _{\mathcal{S}^{d+1}}d^{d+1}\xi_{(d+1)}~\mathcal{L}_{d+1}=\nonumber\\
=\int _{\mathcal{S}^{D+1}}d^{D+1}x \int _{\mathcal{S}^{d+1}} d^{d+1}\xi_{(d+1)}~\delta ^{D+1}
(x^{\mu}-X_{(d+1)}^{\mu}(\xi _{(d+1)}^i))\mathcal{L}_{d+1},
\end{eqnarray}
with $\mathcal{L}_{d+1}$ of the form
\begin{equation}
\mathcal{L}_{d+1}=\epsilon ^{i_1...i_{d+1}}\Omega _{\mu _1...\mu _{d+1}}
\frac{\partial X_{(d+1)}^{\mu _1}}{\partial \xi _{(d+1)}^{i_1}}...
\frac{\partial X_{(2n+1)}^{\mu _{2n+1}}}{\partial \xi _{(d+1)}^{i_{d+1}}},
\end{equation}
where $\epsilon ^{i_1...i_{d+1}}$ is the Levi-Civita symbol, so that the equations of motion 
that correspond to variations of the gauge potentials can be written as a sum of 
various terms describing the coupled dynamic of the gauge fields (that for space-time groups include the vielbein and spin connection) and the brane sources. 

\subsubsection{Case of non zero field strength}

If the field strengths (or curvatures) $F$ and $\overline{F}$ are non zero, the equations of motion corresponding to $\delta A$ are
\begin{equation}
J_A^{\nu (bulk)}+J_A^{\nu (boundary)}+J_A^{\nu (kin)}=0,
\end{equation}
where
\begin{eqnarray}
J_A^{\nu (bulk)}=\sum _{d=2, ~d~even}^{D}\alpha _{d+1}[\frac{d}{2}+1]
\int _{\mathcal{S}^{d+1}}d^{d+1}\xi_{(d+1)}\frac{ \delta ^{D+1}(x^{\nu}-X_{(d+1)}^{\nu}(\xi^{i} _{(d+1)}))}{d!}
\times\nonumber\\
\times <G_AF^{\frac{d}{2}}>_{\mu _2...\mu _{d+1}}
\frac{\partial X_{(d+1)}^{[\nu}}{\partial \xi _{(d+1)}^{i_1}} \frac{\partial X_{(d+1)}^{\mu _2}}{\partial \xi _{(d+1)}^{i_2}}...
\frac{\partial X_{(d+1)}^{\mu _{d+1}]}}{\partial \xi _{(d+1)}^{i_{d+1}}}\epsilon ^{i_1...i_{d+1}}\\
 J_A^{\nu (boundary)}=\sum _{d=2, ~d~even}^{D}\alpha _{d+1}[-d (\frac{d}{2}+1)]
\int _{\mathcal{S}^{d}}d^{d}\xi_{(d)}\frac{\delta ^{D+1}(x^{\nu}-X_{(d)}^{\nu}(\xi^{i} _{(d)}))}{(d-1)!}
\times\nonumber\\
\times \int _0^1dt~ t <\Delta A F_t^{\frac{d}{2}-1}G_A>_{\mu _2...\mu _{d}}
\frac{\partial X_{(d)}^{[\nu}}{\partial \xi _{(d)}^{i_1}} \frac{\partial X_{(d)}^{\mu _2}}{\partial \xi _{(d)}^{i_2}}...
\frac{\partial X_{(d)}^{\mu _{d}]}}{\partial \xi _{(d)}^{i_{d}}}\epsilon ^{i_1...i_{d}}\\
J_A^{\nu (kin)}=\sum _{d=2, ~d~even}^{D}2\beta _{d}
\int _{\mathcal{S}^{d}}d^{d}\xi_{(d)}\delta ^{D+1}(x^{\nu}-X_{(d)}^{\nu}(\xi^{i} _{(d)}))\times\nonumber\\
\times\sqrt{-\gamma _{(d)}} ~\gamma _{(d)}^{ij}<\Delta A_i G_A>
\frac{\partial X_{(d)}^{\nu}}{\partial \xi _{(d)}^{j}}.
\end{eqnarray}

The equations of motion corresponding to $\delta \overline{A}$ are
\begin{equation}
\overline{J}_A^{\nu (bulk)}+\overline{J}_A^{\nu (boundary)}+\overline{J}_A^{\nu (kin)}=0,
\end{equation}
where
\begin{eqnarray}
\overline{J}_A^{\nu (bulk)}=-\sum _{d=2, ~d~even}^{D}\alpha _{d+1}[\frac{d}{2}+1]
\int _{\mathcal{S}^{d+1}}d^{d+1}\xi_{(d+1)}\frac{ \delta ^{D+1}(x^{\nu}-X_{(d+1)}^{\nu}(\xi^{i} _{(d+1)}))}{d!}
\times\nonumber\\
\times <G_A\overline{F}^{\frac{d}{2}}>_{\mu _2...\mu _{d+1}}
\frac{\partial X_{(d+1)}^{[\nu}}{\partial \xi _{(d+1)}^{i_1}} \frac{\partial X_{(d+1)}^{\mu _2}}{\partial \xi _{(d+1)}^{i_2}}...
\frac{\partial X_{(d+1)}^{\mu _{d+1}]}}{\partial \xi _{(d+1)}^{i_{d+1}}}\epsilon ^{i_1...i_{d+1}}\\
 \overline{J}_A^{\nu (boundary)}=\sum _{d=2, ~d~even}^{D}\alpha _{d+1}[-d (\frac{d}{2}+1)]
\int _{\mathcal{S}^{d}}d^{d}\xi_{(d)}\frac{\delta ^{D+1}(x^{\nu}-X_{(d)}^{\nu}(\xi^{i} _{(d)}))}{(d-1)!}
\times\nonumber\\
\times \int _0^1dt~ (1-t) <\Delta A F_t^{\frac{d}{2}-1}G_A>_{\mu _2...\mu _{d}}
\frac{\partial X_{(d)}^{[\nu}}{\partial \xi _{(d)}^{i_1}} \frac{\partial X_{(d)}^{\mu _2}}{\partial \xi _{(d)}^{i_2}}...
\frac{\partial X_{(d)}^{\mu _{d}]}}{\partial \xi _{(d)}^{i_{d}}}\epsilon ^{i_1...i_{d}}\\
\overline{J}_A^{\nu (kin)}=-\sum _{d=2, ~d~even}^{D}2\beta _{d}
\int _{\mathcal{S}^{d}}d^{d}\xi_{(d)}\delta ^{D+1}(x^{\nu}-X_{(d)}^{\nu}(\xi^{i} _{(d)}))\times\nonumber\\
\times\sqrt{-\gamma _{(d)}} ~\gamma _{(d)}^{ij}<\Delta A_i G_A>
\frac{\partial X_{(d)}^{\nu}}{\partial \xi _{(d)}^{j}}.
\end{eqnarray}

\subsubsection{Case of vanishing field strengths}

A somewhat subtle case of the variational problem when the gauge fields are varied occurs if $F=\overline{F}=0$ in the bulk, implying that 
both $A$ and $\overline{A}$ should be pure gauge,
a condition that must hold also at the boundary by continuity. In that case the bulk contribution to the variation of the action vanishes identically, while the variations of the gauge fields at the boundary 
are not the general ones but only those compatible with the pure gauge conditions, which in turn yields weaker equations of motion (or boundary conditions) at the boundary. In that case the equations of motion obtained above for the case in which the gauge fields are not pure gauge must me replaced by the ones resulting from this new variational problem. 

In the pure gauge case $F=\overline{F}=0$ all the bulk equations are satisfied, and only boundary conditions/equations of motion remain. The boundary equations associated to variations of both the metric and the embedding coordinates are also the same, except for the fact that now the gauge field strengths vanish
and the gauge potentials are pure gauge. In the case considered in this subsection $A=g^{-1}dg$ and $\overline{A}=\overline{g}^{-1}d\overline{g}$
locally, with some elements $g$ and $\overline{g}$ of the gauge group. 
In this case $\delta A=D(g^{-1}\delta g)=d(g^{-1}\delta g)+(g^{-1}d g)(g^{-1}\delta g)-(g^{-1}\delta g)(g^{-1}d g)$ and $\delta \overline{A}=\overline{D}(\overline{g}^{-1}\delta \overline{g})=
d(\overline{g}^{-1}\delta \overline{g})+(\overline{g}^{-1}d \overline{g})(\overline{g}^{-1}\delta \overline{g})-(\overline{g}^{-1}\delta \overline{g})(\overline{g}^{-1}d \overline{g})$. The required variations of the various terms of the action are, obtained integrating by parts and discarding total derivatives are
\begin{eqnarray}
\delta _A S_d^{(Kin)}=2\beta _d\int _{\mathcal{S}^{d}} d^d\xi \sqrt{-\gamma _d}\gamma _d^{ij}
<\mathcal{D}_j(\Delta A_i)(g^{-1}\delta g)>\\
\delta _{\overline{A}} S_d^{(Kin)}=-2\beta _d\int _{\mathcal{S}^{d}} d^d\xi \sqrt{-\gamma _d}\gamma _d^{ij}
<\overline{\mathcal{D}}_j(\Delta A_i)(\overline{g}^{-1}\delta \overline{g})>,
\end{eqnarray}
where the gauge and world-volume covariant derivatives are 
$\mathcal{D}_i(\Delta A_j)=\partial _i(\Delta A_j)+[A_i,\Delta A_j]-\Gamma ^k_{ij}\Delta A_k$, 
with $\Gamma ^k_{ij}$ the Christoffel symbol corresponding to the world-volume metrics $\gamma _{ij}$, and
\begin{eqnarray}
\delta _A S_d^{(Trans)}=-\alpha _d \frac{d}{2}(\frac{d}{2}+1)(d-1)B(1+\frac{d}{2},\frac{d}{2})\int _{\mathcal{S}^{d}} 
<(\Delta A)^d(g^{-1}\delta g)>\\
\delta _{\overline{A}} S_d^{(Trans)}=+\alpha _d \frac{d}{2}(\frac{d}{2}+1)(d-1)B(\frac{d}{2},1+\frac{d}{2})\int _{\mathcal{S}^{d}} 
<(\Delta \overline{A})^d(\overline{g}^{-1}\delta \overline{g})>,
\end{eqnarray}
where $B(1+\frac{d}{2},\frac{d}{2})=B(\frac{d}{2},1+\frac{d}{2})$ es the Euler Beta function of the given arguments. With these variations it is straightforward to proceed as in the previous subsection, using Dirac deltas and factorizing the $g^{-1}\delta g$ and $\overline{g}^{-1}\delta \overline{g}$ variational parameter to leave group generators $G_A$, to obtain the equations of motion of the branes interacting with the fields for which they are also sources. These equations have now the generic form
$\mathcal{J}_A^{\nu (boundary)}+\mathcal{J}_A^{\nu (kin)}=0$, without the bulk contribution and with the currents implied by the variations given above.\\

\subsection{Equations of motion associated to the variation of the world volume metric}

The world-volume metric $\gamma_{(d)kr}$ appears in the action only in the kinetic term $S_{d}^{(Kin)}$.
Its equations of motion are
\begin{eqnarray}
<\Delta A_i\Delta A_j>-\frac{1}{2}\gamma _{(d)ij}[\gamma _{(d)}^{kl}<\Delta A_k\Delta A_l>-(d-2)]=0.
\end{eqnarray}
The solution of these equations is
\begin{equation}
\gamma _{(d)ij}=<\Delta A_i\Delta A_j>.
\end{equation}
In $d=2$, just like in the case of the Polyakov String Action, these equations of motion only imply a local proportionality between both members of the previous equation, but in that dimension the action is invariant under local rescaling of the metric,  allowing to stablish the equality also in that case.

\subsection{Equations of motion associated to the variation of the embedding coordinates of the branes}

To determine the equations of motion associated to the $X_{(d+1)}^{\mu}$ we need that the variation of $S^{(Trans)}_{d+1}$ and $S_{d}^{(Kin)}$ under a variation of $X_{(d)}^{\mu}$  are
\begin{eqnarray}
\delta _X S^{(Trans)}_{d+1}=\alpha _d \int _{\mathcal{S}^{d+1}}d^{d+1}\xi _{(d+1)}
\mathfrak{F}_{(d+2)\nu\mu _1...\mu_{d+1}}\frac{\partial X^{\mu _1}_{(d+1)}}{\partial \xi^{i_1} _{(d+1)}}...
\frac{\partial X^{\mu _{d+1}}_{(d+1)}}{\partial \xi^{i_{d+1}} _{(d+1)}}\epsilon ^{i_1...i_{d+1}}
\delta X_{(d+1)}^{\nu}+\nonumber\\
+\alpha _d \int _{\mathcal{S}^{d+1}}d^{d+1}\xi _{(d+1)}\frac{\partial ~}{\partial \xi _{(d+1)}^k}
\left[ \mathcal{P}_{(d+1)\nu }^k\delta X_{(d+1)}^{\nu}\right]\\
\delta _X S_{d}^{(Kin)}=2\beta _d\int _{\mathcal{S}^d=
\partial\mathcal{S}^{d+1}}d^d\xi _{d}\times\nonumber\\
\times\left[  \sqrt{-\gamma _{(d)}}\gamma _{(d)}^{ij}<\Delta A_i\frac{\partial \Delta A_{\mu}}{\partial X _{(d)}^{\nu }}>  
\frac{\partial X^{\mu }_{(d)}}{\partial \xi^{j} _{(d)}} -\frac{\partial ~}{\partial \xi^{j} _{(d)}}
\left(\sqrt{-\gamma}\gamma ^{ij}_{(d)}<\Delta A_i\Delta A_{\nu }>\right)
\right]\delta X_{(d)}^{\nu}+\nonumber\\
+ 2\beta _d\int _{\mathcal{S}^d=
\partial\mathcal{S}^{d+1}}d^d\xi _{d} \frac{\partial ~}{\partial \xi^{j} _{(d)}}
\left(\sqrt{-\gamma}\gamma ^{ij}_{(d)}<\Delta A_i\Delta A_{\nu }>\delta X_{(d)}^{\nu}\right),
\end{eqnarray}
where $\mathfrak{F}_{(d+2)}$ and the momenta conjugated to $X_{(d+1)}^{\nu}$,  $\mathcal{P}_{(d+1)\nu }^k$, are
\begin{eqnarray}
\mathfrak{F}_{(d+2)}\equiv d\mathfrak{T}_{d+1}=<F^{\frac{d}{2}+1}>-<\overline{F}^{\frac{d}{2}+1}>\\
\mathcal{P}_{(d+1)\nu }^k=\frac{1}{d!}[\mathfrak{T}_{d+1}]_{\nu\mu _2...\mu _{d+1}}
\frac{\partial X^{\mu _2}_{(d+1)}}{\partial \xi^{i_2} _{(d+1)}}...
\frac{\partial X^{\mu _{d+1}}_{(d+1)}}{\partial \xi^{i_{d+1}} _{(d+1)}}\epsilon ^{k i_2...i_{d+1}}.
\end{eqnarray}
Considering that $X_{(d+1)}$ and $X_{(d)}$ are the same at the same point of the boundary of the brane,
we obtain the equations of motion for the bulk
\begin{equation}
\mathfrak{F}_{(d+2)\nu\mu _1...\mu_{d+1}}\frac{\partial X^{\mu _1}_{(d+1)}}{\partial \xi^{i_1} _{(d+1)}}...
\frac{\partial X^{\mu _{d+1}}_{(d+1)}}{\partial \xi^{i_{d+1}} _{(d+1)}}\epsilon ^{i_1...i_{d+1}}=0.
\end{equation}
At the boundary, for those values of $\nu$ with $X^{\nu}$ corresponding to free boundary conditions we obtain the conditions/equations of motion
\begin{eqnarray}
0=\alpha _d  ~  \mathcal{P}_{(d+1)\nu }^k n_{(d)k}                          +\nonumber\\
+2\beta _d \left[  \sqrt{-\gamma _{(d)}}\gamma _{(d)}^{ij}<\Delta A_i\frac{\partial \Delta A_{\mu}}{\partial X _{(d)}^{\nu }}>  
\frac{\partial X^{\mu }_{(d)}}{\partial \xi^{j} _{(d)}} -\frac{\partial ~}{\partial \xi^{j} _{(d)}}
\left(\sqrt{-\gamma}\gamma ^{ij}_{(d)}<\Delta A_i\Delta A_{\nu }>\right)
\right],
\end{eqnarray}
where $n_{(d)k}$ is the normal to the boundary from Gauss's Theorem (see Ref.\cite{Schutz}, section 4.23). For instance, if the boundary coordinates are such that $\xi^{j} _{(d)}=\xi^{j} _{(d+1)}$ if $j=0,...,d-1$,
then $n_{(d)k}=\delta _{kd}$. No further equations are obtained for those values of $\nu$ such that Dirichlet boundary conditions 
$\delta X^{\nu}=0$ are imposed. 

\section{Comments on form of the Action, Symmetries and Equations of Motion}

\subsection{Symmetries}

The symmetries of the Chern-Simons action were discussed in the original article\cite{Mora-Nishino20001}. These are: invariance under general coordinate transformations of the background
space-time and the branes world volumes, and gauge invariance under gauge transformations that are the same for both $A$ and $\overline{A}$, given by $A^h=h^{-1}[A+d]h$ and 
$\overline{A}^h=h^{-1}[\overline{A}+d]h$ for any element $h$ of the gauge group. Gauge invariance 
follows from the properties of the symmetrized trace and the covariance of $F$, $\overline{F}$, $F_t$
and $\Delta A $.

The action is not invariant under gauge transformations for which only one of the gauge fields transforms, or both of them transform with different gauge group elements, but these variations yield only boundary terms, therefore the action is said to be {\it quasi-invariant} under these transformations. 
It is worthwhile to emphasize that possible bulk terms of the form $<(\Delta A)^{d+1}>$ are invariant under gauge transformations that transform both gauge fields in the same way, but they are neither
invariant nor quasi-invariant under gauge transformations of only one of the gauge fields. 
As $d$ is even, candidate boundary terms $<(\Delta A)^{d}>$ are in fact identically zero. The action of Chern-Simons branes satisfies then not only the condition of being gauge invariant under transformations of both gauge potentials with the same group element, but also the above mentioned quasi-invariance condition, which can be taken as a requirement on the form of the action that rules out terms such as $<(\Delta A)^{d+1}>$. It is also true, and related with the previous remarks, that if both gauge potentials are pure gauge, then the only local degrees of freedom that remain are located at the boundaries, and there are no local degrees of freedom in the bulk, what can be seen as a holographic property of the action.

The interchange of both gauge potentials transforms the transgression part of the action by a minus sign, while it leaves the kinetic part unchanged.  

\subsection{Special choices for the Action}

The requirement on the form of the action mentioned in the previous subsection, that it be such that in the pure gauge case only boundary degrees of freedom remain, and the related condition that under gauge transformations of only one of the gauge potentials the action must be quasi-invariant, exclude not only terms of the form $<(\Delta A)^{d+1}>$, but also kinetic terms with support in the bulk. 

Furthermore, if one imposes the condition that the only true degrees of freedom are those of the gauge fields one must exclude also all the boundary kinetic terms except for the $d=2$ one, making $\beta _d=0$
for $d\neq 2$. The reason is that in $d=2$, because of Weyl invariance, the metric is an auxiliary variable with no local degrees of freedom, which can be locally reduced to the flat metric, while that is not the case for $d\neq 2$, and one should in principle add also terms involving the curvature of these metrics.  
In $d=2$ there are two choices of $\beta _2$ that result in special properties for the action, particularly  the decoupling of left and right movers, and Weyl invariance of the action, in a way analogous to Ref.\cite{Witten-bosonization}. It is interesting that the same relative coefficient between the kinetic and transgression parts resulted in Ref.\cite{Mora-Nishino20001} from the requirement of kappa symmetry.

We define 
\begin{eqnarray}
\mathbb{P}^{ij}_{\pm}\equiv\frac{1}{2}[\gamma ^{ij}\pm \frac{\epsilon ^{ij}}{\sqrt{-\gamma}}],
\end{eqnarray}
wich has the properties of a projector: 
$\mathbb{P}^{ik}_{\pm}\mathbb{P}^{~j}_{k\pm}=\mathbb{P}^{ij}_{\pm}$ 
(that is, $\mathbb{P}^{2}_{\pm}=\mathbb{P}_{\pm}$) and 
$\mathbb{P}^{ik}_{\pm}\mathbb{P}^{~j}_{k\mp}=\mathbb{P}^{ik}_{\mp}\mathbb{P}^{~j}_{k\pm}=0$ 
(that is, $\mathbb{P}_{\pm}\mathbb{P}_{\mp}=\mathbb{P}_{\mp}\mathbb{P}_{\pm}=0$ ). Furthermore 
$\mathbb{P}^{ij}_{\pm}=\mathbb{P}^{ji}_{\mp}$, or $\mathbb{P}^{T}_{\pm}=\mathbb{P}_{\mp}$. Given vectors $V_i$ and $W_j$ in $\mathcal{S}^2$ we define the projections $V^i_{\pm}=\mathbb{P}^{ij}_{\pm}V_j$, 
$W^i_{\pm}=\mathbb{P}^{ij}_{\pm}W_j$, and 
it holds that $\mathbb{P}^{ij}_{\pm}W_iV_j=W_{\mp i}V^i_{\pm}$.

There are two special choices for $\beta _2$, which are $\beta _2=\pm\frac{1}{2}\alpha _2$. For these choices the variation of the $d=2$ part of the action under variations of $A$ and $\overline{A}$ are
\begin{eqnarray}
\delta _A[S_2^{(Kin)}+S_3^{(Trans)}]=2\alpha _2\int_{\mathcal{S}^3} <F\delta A>+
2\alpha _2\int _{\mathcal{S}^2=\partial \mathcal{S}^3}d^2\xi \sqrt{-\gamma}~ 
\mathbb{P}^{ij}_{\pm}<\Delta A_i\delta A_j>=\nonumber\\
=2\alpha _2\int_{\mathcal{S}^3} <F\delta A>+
2\alpha _2\int _{\mathcal{S}^2=\partial \mathcal{S}^3}d^2\xi \sqrt{-\gamma}~<\Delta A_{\mp i}\delta A_{\pm}^i>\\
\delta _{\overline{A}}[S_2^{(Kin)}+S_3^{(Trans)}]=-2\alpha _2\int_{\mathcal{S}^3} <\overline{F}\delta \overline{A}>-
2\alpha _2\int _{\mathcal{S}^2=\partial \mathcal{S}^3}d^2\xi \sqrt{-\gamma}~ 
\mathbb{P}^{ij}_{\mp}<\Delta A_i\delta \overline{A}_j>=\nonumber\\
=-2\alpha _2\int_{\mathcal{S}^3} <\overline{F}\delta \overline{A}>-
2\alpha _2\int _{\mathcal{S}^2=\partial \mathcal{S}^3}d^2\xi \sqrt{-\gamma}~<\Delta A_{\pm i}\delta \overline{A}_{\mp}^i>.
\end{eqnarray}
In the pure gauge case of vanishing field strengths we have
\begin{eqnarray}
\delta _A[S_2^{(Kin)}+S_3^{(Trans)}]=\pm \alpha _2\int _{\mathcal{S}^{2}} d^2\xi
~\mathbb{P}^{ij}_{\pm}<\mathcal{D}_j(\Delta A_i)(g^{-1}\delta g)>=\nonumber\\
=\pm \alpha _2\int _{\mathcal{S}^{2}} d^2\xi
<\mathcal{D}_{\pm }(\Delta A_{\mp })(g^{-1}\delta g)>\\
\delta _{\overline{A}}[S_2^{(Kin)}+S_3^{(Trans)}]=-\pm\alpha _2\int _{\mathcal{S}^{2}} d^2\xi ~\mathbb{P}^{ij}_{\pm}
<\overline{\mathcal{D}}_j(\Delta A_i)(\overline{g}^{-1}\delta \overline{g})>=\nonumber\\
=-\pm\alpha _2\int _{\mathcal{S}^{2}} d^2\xi 
<\overline{\mathcal{D}}_{\pm}(\Delta A_{\mp})(\overline{g}^{-1}\delta \overline{g})>.
\end{eqnarray}
For instance, the equation of motion obtained from eq.(32), if there are no other overlapping branes, is 
$<\mathcal{D}_{\pm }(\Delta A_{\mp })G_A>=0$, but Weyl invariance allows to choose the world-volume metric to be flat, what  makes the Christoffel symbols zero, and $\mathcal{D}_{\pm }=D _{\pm }$ and 
$\mathbb{P}^{ij}_{\pm}\equiv\frac{1}{2}[\eta ^{ij}\pm \epsilon ^{ij}]$. In that case 
$0=<\mathcal{D}_{\pm }(\Delta A_{\mp })G_A>=<D_{\pm }(\Delta A_{\mp })G_A>=
\partial _{\pm }(\Delta A_{\mp })_B<G_B G_A>$. If we take $\overline{A}=0$ and $<G_B G_A>$ is invertible, then we obtain 
$\partial _{\pm }(g^{-1}\partial_{\mp }g)_B=0$, as in Ref.\cite{Witten-bosonization}, which implies the decoupling of right and left movers, as claimed.

 \section{The Bosonic String}

The AdS group in $D+1$ dimensions $SO(D-1,2)$, which is also the Conformal group in $D$ dimensions, corresponds to linear transformations that leave invariant
the metric $\eta _{AB}$, $A,B=0,...,D+1$, with zero off-diagonal entries and diagonal elements $(-1,+1,+1,...,+1,-1)$. The algebra of the generators $J_{AB}$ of the AdS group is
\begin{eqnarray}
[J_{AB},J_{CD}]=+\eta _{AC} J_{BD} +\eta _{BD} J_{AC} -\eta _{BC}J_{AD} -\eta _{AD} J _{BC} 
\end{eqnarray}
Splitting the indices into $a,b=0,...,d$ and $D+1\equiv *$, and defining the 
Lorentz generators $J_{ab}$ and the linear momentum generator $P_a=\frac{J_{a*}}{l}$, where $l$ is a real and positive parameter with the dimension of length called tha "AdS radius", we get
\begin{eqnarray}
[J_{ab},J_{cd}]=+\eta _{ac} J_{bd} +\eta _{bd} J_{ac} -\eta _{bc}J_{ad} -\eta _{ad} J _{bc}\\
\left[ J_{ab},P_{c}\right] =+\eta _{ac} P_{b} + -\eta _{bc}P_{a}\\
\left[ P_{a},P_{b}\right] =+\frac{1}{l^2}\eta _{**} J_{ab}=-\frac{1}{l^2} J_{ab}
\end{eqnarray}
In la limit $l\rightarrow\infty$, known as Inonu-Wigner contraction, 
this algebra turns into the Poincar\'e group algebra in $D+1$ dimensions.

The invariant trace of two generators is
\begin{eqnarray}
STr[J_{ab}J_{cd}]=\eta _{ac} \eta _{bd} -\eta _{ad} \eta _{bc}\\
Str[J_{ab}P_c]=0~~~,~~~STr[P_aP_b]=-\frac{1}{l^2}\eta _{ab}
\end{eqnarray}

It is possible to show that the bosonic string with Polyakov Action is obtained as part of the space of solutions of Chern-Simons branes in a background Chern-Simons AdS gravity in the limit of large $l$, by proceeding as follows.

We consider gauge group elements of the form $g(x)=\exp[-\mathcal{V} ^a(x)P_a]$, where the gauge parameters $\mathcal{V} ^a(x)$ are some function of the space-time point, instead of the generic
$g(x)=\exp[-\mathcal{V} ^a(x)P_a+\mathcal{W}^{ab}(x)J_{ab}]$ . 
Notice that $\mathcal{V} ^a P_a=\frac{\mathcal{V} ^a}{l}J_{a*}$, where the $J_{a*}$ are dimensionless, and therefore the $\mathcal{V} ^a$ must have the dimension of length. Assuming the exponent to be small, which amounts to require $\mathcal{V} ^a\ll l$, we can consider only the first terms of the expansions for the group element $g=I+\mathcal{V} ^aP_a+\frac{1}{2}\mathcal{V} ^a\mathcal{V} ^bP_aP_b+...$, its inverse
$g^{-1}=I-\mathcal{V} ^aP_a+\frac{1}{2}\mathcal{V} ^a\mathcal{V} ^bP_aP_b+...$, 
its exterior derivative
$dg=d\mathcal{V} ^aP_a+\frac{1}{2}d\mathcal{V} ^a\mathcal{V} ^bP_aP_b+\frac{1}{2}\mathcal{V} ^ad\mathcal{V} ^bP_aP_b+...$ and the Maurer-Cartan form 
$g^{-1}dg=d\mathcal{V} ^aP_a+\frac{1}{2 l^2}d\mathcal{V} ^a\mathcal{V} ^bJ_{ab}+...$, with leading order $\mathcal{V}/l$. 
Considering the form of the gauge potential in Chern-Simons AdS gravity in terms of the 
vielbein $e^a=e^a_{\mu}dx^{\mu}$ and the spin connection 
$\omega ^{ab}=\omega ^{ab}_{\mu}dx^{\mu}$, $A=e^aP_a+\frac{\omega ^{ab}}{2}J_{ab}$, then the pure gauge potential associated to the Maurer-Cartan form $A^{(pure~gauge)}=g^{-1}dg$ implies 
$e^a=d\mathcal{V}^a$ and $\omega ^{ab}=-\frac{1}{4l^2}\mathcal{V}^{[a}d\mathcal{V}^{b]}$. Notice that the vielbein part is the one of leading order in the  expansion in powers of $\mathcal{V}^a/l$, even considering, as we must, $e^a/l$.
We make a choice of coordinates such that $\frac{\partial \mathcal{V}^a}{\partial x^{\mu}}=\delta ^ a_{\mu}+...$, or $\mathcal{V}^a=\delta ^ a_{\mu}x^{\mu}+...$, valid in a neighbourhood of $x^{\mu}=0$ such that $x^{\mu}\ll l$. For this coordinate choice the vielbein is $e^a_{\mu}=\delta^a_{\mu}+...$.

Given the embeddings $X ^{\mu }(\xi ^i )$ we obtain the pull-backs 
$\partial _i\mathcal{V}^a=\frac{\partial\mathcal{V}^a}{\partial \xi ^i}=\frac{\partial\mathcal{V}^a}{\partial X^{\mu}}\frac{\partial X^{\mu}}{\partial \xi ^i}=\frac{\partial\mathcal{V}^a}{\partial X^{\mu}}\partial _i X^{\mu}$. With the approximations and choice of coordinates considered $\partial _i\mathcal{V}^a=\delta ^a_i\partial _i X^{\mu}+...=\partial _i X^{a}+...$.

We focus now in the case of $F=0$ ($A$ pure gauge), concretely an AdS space-time, and a 2-brane with Dirichlet boundary conditions and its boundary in the flat AdS boundary. For the $\overline{A}$ configuration we choose the "Kounterterms" vacuum of Ref.\cite{Mora20141}, 
that has a zero vielbein $\overline{e}=0$, and a non-zero field strength $\overline{F}\neq 0$, but does nonetheless satisfy the bulk field equations.
In that case, and keeping only the leading order terms in $1/l$, the bulk equations are satisfied for the bulk of both the background and any branes, and the only remaining boundary contribution corresponds to the boundary of the 2-branes.  
We have 
$STr[(g^{-1}\partial _i g)(g^{-1}\partial _j g)]=-\frac{1}{l^2}\eta _{ab}\partial _i X^a\partial _jX^b+...$. This means that $STr[A\overline{A}]=0$ to that order, as it is proportional to the previous expression contracted with antisymmetric $\epsilon ^{ij}$. The WZW term of the action has contributions of order higher than $1/l^3$, and what remains of the 2-brane action is the Polyakov action
\begin{eqnarray}
S^{(Kin)}_2= -\beta _2 \int _{\mathcal{S}^2}d^2\xi\sqrt{-\gamma }\gamma ^{ij}\frac{1}{l^2}\eta _{ab}\partial _i X^a\partial _jX^b+...
\end{eqnarray}
Notice that the coordinate normal to the boundary is fixed for the brane $\partial _iX^{D}=0$ and hence in the previous expression $a,b=0,...,D-1$. 

\section{Quantization of the constants in the Brane Actions}

There are at least two ways to see that the constants in front of the brane actions (tensions or coupling constants) must be quantized.

The first argument, which was originally given in Refs.\cite{Deser:1981wh,Witten:1983tw,Witten-bosonization}, applies to branes with boundaries in the pure gauge ($F=0$) case, where the actions reduce to WZW actions at the boundary, arises from the condition that the Feynman Path Integral must be invariant under large gauge transformations and independent of the way in which the boundary is extended into the bulk to define te WZW term. 
The existence of non equivalent ways of extending the branes into de bulk is classified 
by the homotopy groups $\pi _{d+1}(\mathcal{G})$ for the d-brane. 
The space-time group $\mathcal{G}$ is en general non compact, then the relevant result is that its homotopy groups are those of its maximal compact subgroups. For instance, if $\mathcal{G}$ is the AdS group $SO(d-1,2)$, then its maximal compact subgroup is $SO(d-1)\times SO(2)$, 
and $\pi _{d+1}(SO(d-1,2))=\pi _{d+1}(SO(d-1)\times SO(2))=
\pi _{d+1}(SO(d-1))\oplus \pi _{d+1}(SO(2))= \pi _{d+1}(SO(d-1))$, which can be looked up in standard tables. In the case of the 2-brane the relevant homotopy group is $\pi _3(SO(d-1))=Z$, implying that its tension constant is quantized.  

A second way to see that the tensions are quantized, valid for closed branes, is analogous to the Dirac quantization condition, and also to the argument of Ref.\cite{Zanelli:1994ti}. The idea is that if one swings a d-brane in a loop (a d+1 dimensional manifold), and then shrinks that loop until it vanishes, so that the brane actually does not move, then the amplitude for this process must be 1, but that amplitude is $\exp{(iS/\hbar)}$, which when the loop shrinks to zero is $\exp{(\frac{i}{\hbar}\int _{M^{d+2}}<F^{\frac{d}{2}+1}>)}$ (where $M^{d+2}$ is some manifold with the loop as a boundary, and without boundary when the loop vanishes) and must be $e^{i 2\pi N}$ for some integer N. It results $\int _{M^{d+2}}<F^{\frac{d}{2}+1}>=2\pi\hbar N$, which is possible if the symmetrized trace $<...>$ is such that the integral corresponds to some index theorem, and the tension is quantized. If the integral considered does not vanish that would signal de presence of a dual "magnetic" solitonic (D-d-3)-brane. Notice that these conditions constrain the set of acceptable symmetrized traces.

\section{Conclusions and further directions}

The class of theories considered in this Letter has several attractive properties that merit further investigation.
Among possible lines for future research are:
\begin{enumerate}[(i)]

\item The detailed study of supersymmetric cases in which the gauge group is a supergroup, which was started in Ref.\cite{Mora-Nishino20001}, with the goal of finding a model with a suitable symmetrised invariant trace and gauge group to connect with various known limits of M-theory.

\item The analysis of conserved charges and anomalies of these theories.

\item Looking for solitonic dual "magnetic" brane solutions to these theories, 
in analogy to \cite{Duff:1993ye}, and looking for dualities between fundamental and solitonic branes, reinterpreting the strategy of Refs.\cite{Dixon:1992if,Duff:1994vv,Duff:1995wd} in our framework.
In those references it is also pointed out that duality mixes tree level (classical) effects with one-loop (quantum) effects. It would be interesting to know if that has a counterpart in our framework, and what is its meaning if it does.
 
\end{enumerate}

\bibliography{mora-V1995-bibfyle}

\begin{thebibliography}{10}
\expandafter\ifx\csname url\endcsname\relax
  \def\url#1{\texttt{#1}}\fi
\expandafter\ifx\csname urlprefix\endcsname\relax\def\urlprefix{URL }\fi
\expandafter\ifx\csname href\endcsname\relax
  \def\href#1#2{#2} \def\path#1{#1}\fi

\bibitem{Mora-Nishino20001}
P.~Mora, H.~Nishino, {Fundamental extended objects for Chern–Simons
  supergravity}, Physics Letters B 482 (2000) 222--232.
\newblock \href {http://arxiv.org/abs/hep-th/0002077}
  {\path{arXiv:hep-th/0002077}}, \href
  {https://doi.org/10.1016/S0370-2693(00)00535-9}
  {\path{doi:10.1016/S0370-2693(00)00535-9}}.

\bibitem{Mora20011}
P.~Mora, {Chern–Simons supersymmetric branes}, Nuclear Physics B 594 (2001)
  229--242.
\newblock \href {http://arxiv.org/abs/hep-th/0008180}
  {\path{arXiv:hep-th/0008180}}, \href
  {https://doi.org/10.1016/S0550-3213(00)00649-0}
  {\path{doi:10.1016/S0550-3213(00)00649-0}}.

\bibitem{Achucarro:1987vz}
A.~Achucarro, P.~Townsend, {A Chern-Simons Action for Three-Dimensional anti-De
  Sitter Supergravity Theories}, Phys. Lett. B 180 (1986) 89.
\newblock \href {https://doi.org/10.1016/0370-2693(86)90140-1}
  {\path{doi:10.1016/0370-2693(86)90140-1}}.

\bibitem{Chamseddine:1990gk}
A.~H. Chamseddine, {Topological gravity and supergravity in various
  dimensions}, Nucl. Phys. B 346 (1990) 213--234.
\newblock \href {https://doi.org/10.1016/0550-3213(90)90245-9}
  {\path{doi:10.1016/0550-3213(90)90245-9}}.

\bibitem{Chamseddine:1989nu}
A.~H. Chamseddine, {Topological Gauge Theory of Gravity in Five-dimensions and
  All Odd Dimensions}, Phys. Lett. B 233 (1989) 291--294.
\newblock \href {https://doi.org/10.1016/0370-2693(89)91312-9}
  {\path{doi:10.1016/0370-2693(89)91312-9}}.

\bibitem{MuellerHoissen:1990vf}
F.~Mueller-Hoissen, {From Chern-Simons to Gauss-Bonnet}, Nucl. Phys. B 346
  (1990) 235--252.
\newblock \href {https://doi.org/10.1016/0550-3213(90)90246-A}
  {\path{doi:10.1016/0550-3213(90)90246-A}}.

\bibitem{Banados:1993ur}
M.~Banados, C.~Teitelboim, J.~Zanelli, {Dimensionally continued black holes},
  Phys. Rev. D 49 (1994) 975--986.
\newblock \href {http://arxiv.org/abs/gr-qc/9307033}
  {\path{arXiv:gr-qc/9307033}}, \href {https://doi.org/10.1103/PhysRevD.49.975}
  {\path{doi:10.1103/PhysRevD.49.975}}.

\bibitem{Banados:1996hi}
M.~Banados, R.~Troncoso, J.~Zanelli, {Higher dimensional Chern-Simons
  supergravity}, Phys. Rev. D 54 (1996) 2605--2611.
\newblock \href {http://arxiv.org/abs/gr-qc/9601003}
  {\path{arXiv:gr-qc/9601003}}, \href
  {https://doi.org/10.1103/PhysRevD.54.2605}
  {\path{doi:10.1103/PhysRevD.54.2605}}.

\bibitem{Zanelli:2012px}
J.~Zanelli, {Chern-Simons Forms in Gravitation Theories}, Class. Quant. Grav.
  29 (2012) 133001.
\newblock \href {http://arxiv.org/abs/1208.3353} {\path{arXiv:1208.3353}},
  \href {https://doi.org/10.1088/0264-9381/29/13/133001}
  {\path{doi:10.1088/0264-9381/29/13/133001}}.

\bibitem{Dixon:1991xz}
J.~A. Dixon, M.~Duff, E.~Sezgin, {The Coupling of Yang-Mills to extended
  objects}, Phys. Lett. B 279 (1992) 265--271.
\newblock \href {http://arxiv.org/abs/hep-th/9201019}
  {\path{arXiv:hep-th/9201019}}, \href
  {https://doi.org/10.1016/0370-2693(92)90391-G}
  {\path{doi:10.1016/0370-2693(92)90391-G}}.

\bibitem{Green:1989nn}
M.~B. Green, {Supertranslations, Superstrings and \{Chern-Simons\} Forms},
  Phys. Lett. B 223 (1989) 157--164.
\newblock \href {https://doi.org/10.1016/0370-2693(89)90233-5}
  {\path{doi:10.1016/0370-2693(89)90233-5}}.

\bibitem{Troncoso1997va}
R.~Troncoso, J.~Zanelli, {New gauge supergravity in seven-dimensions and
  eleven-dimensions}, Phys. Rev. D 58 (1998) 101703.
\newblock \href {http://arxiv.org/abs/hep-th/9710180}
  {\path{arXiv:hep-th/9710180}}, \href
  {https://doi.org/10.1103/PhysRevD.58.101703}
  {\path{doi:10.1103/PhysRevD.58.101703}}.

\bibitem{Troncoso1998ng}
R.~Troncoso, J.~Zanelli, {Gauge supergravities for all odd dimensions}, Int. J.
  Theor. Phys. 38 (1999) 1181--1206.
\newblock \href {http://arxiv.org/abs/hep-th/9807029}
  {\path{arXiv:hep-th/9807029}}, \href
  {https://doi.org/10.1023/A:1026614631617}
  {\path{doi:10.1023/A:1026614631617}}.

\bibitem{Cremmer:1978km}
E.~Cremmer, B.~Julia, J.~Scherk, {Supergravity Theory in Eleven-Dimensions},
  Phys. Lett. B 76 (1978) 409--412.
\newblock \href {https://doi.org/10.1016/0370-2693(78)90894-8}
  {\path{doi:10.1016/0370-2693(78)90894-8}}.

\bibitem{Horava:1997dd}
P.~Horava, {M theory as a holographic field theory}, Phys. Rev. D 59 (1999)
  046004.
\newblock \href {http://arxiv.org/abs/hep-th/9712130}
  {\path{arXiv:hep-th/9712130}}, \href
  {https://doi.org/10.1103/PhysRevD.59.046004}
  {\path{doi:10.1103/PhysRevD.59.046004}}.

\bibitem{Hull:1994ys}
C.~Hull, P.~Townsend, {Unity of superstring dualities}, Nucl. Phys. B 438
  (1995) 109--137.
\newblock \href {http://arxiv.org/abs/hep-th/9410167}
  {\path{arXiv:hep-th/9410167}}, \href
  {https://doi.org/10.1016/0550-3213(94)00559-W}
  {\path{doi:10.1016/0550-3213(94)00559-W}}.

\bibitem{Witten:1995ex}
E.~Witten, {String theory dynamics in various dimensions}, Nucl. Phys. B 443
  (1995) 85--126.
\newblock \href {http://arxiv.org/abs/hep-th/9503124}
  {\path{arXiv:hep-th/9503124}}, \href
  {https://doi.org/10.1016/0550-3213(95)00158-O}
  {\path{doi:10.1016/0550-3213(95)00158-O}}.

\bibitem{Duff:1995wd}
M.~J. Duff, J.~T. Liu, R.~Minasian, {Eleven-dimensional origin of string-string
  duality: A One loop test}, Nucl. Phys. B 452 (1995) 261--282.
\newblock \href {http://arxiv.org/abs/hep-th/9506126}
  {\path{arXiv:hep-th/9506126}}, \href
  {https://doi.org/10.1016/0550-3213(95)00368-3}
  {\path{doi:10.1016/0550-3213(95)00368-3}}.

\bibitem{Edelstein:2008ry}
J.~D. Edelstein, J.~Zanelli, {Sources for Chern-Simons theories}, in: {workshop
  on Quantum Mechanics of Fundamental Systems: the Quest for Beauty and
  Simplicity: Dedicated to Claudio Bunster on the occasion of his 60th
  birthday}, 2009, pp. 107--124.
\newblock \href {http://arxiv.org/abs/0807.4217} {\path{arXiv:0807.4217}},
  \href {https://doi.org/10.1007/978-0-387-87499-9_8}
  {\path{doi:10.1007/978-0-387-87499-9_8}}.

\bibitem{Miskovic:2009dd}
O.~Miskovic, J.~Zanelli, {Couplings between Chern-Simons gravities and
  2p-branes}, Phys. Rev. D 80 (2009) 044003.
\newblock \href {http://arxiv.org/abs/0901.0737} {\path{arXiv:0901.0737}},
  \href {https://doi.org/10.1103/PhysRevD.80.044003}
  {\path{doi:10.1103/PhysRevD.80.044003}}.

\bibitem{Edelstein:2010sx}
J.~D. Edelstein, A.~Garbarz, O.~Miskovic, J.~Zanelli, {Stable p-branes in
  Chern-Simons AdS supergravities}, Phys. Rev. D 82 (2010) 044053.
\newblock \href {http://arxiv.org/abs/1006.3753} {\path{arXiv:1006.3753}},
  \href {https://doi.org/10.1103/PhysRevD.82.044053}
  {\path{doi:10.1103/PhysRevD.82.044053}}.

\bibitem{Edelstein:2010sh}
J.~D. Edelstein, A.~Garbarz, O.~Miskovic, J.~Zanelli, {Naked Singularities,
  Topological Defects and Brane Couplings}, Int. J. Mod. Phys. D 20 (2011)
  839--849.
\newblock \href {http://arxiv.org/abs/1009.4418} {\path{arXiv:1009.4418}},
  \href {https://doi.org/10.1142/S0218271811019177}
  {\path{doi:10.1142/S0218271811019177}}.

\bibitem{Edelstein:2011vu}
J.~D. Edelstein, A.~Garbarz, O.~Miskovic, J.~Zanelli, {Geometry and stability
  of spinning branes in AdS gravity}, Phys. Rev. D 84 (2011) 104046.
\newblock \href {http://arxiv.org/abs/1108.3523} {\path{arXiv:1108.3523}},
  \href {https://doi.org/10.1103/PhysRevD.84.104046}
  {\path{doi:10.1103/PhysRevD.84.104046}}.

\bibitem{Kastikainen:2020auf}
J.~Kastikainen, {Conical defects and holography in topological AdS gravity},
  Class. Quant. Grav. 37~(19) (2020) 195010.
\newblock \href {http://arxiv.org/abs/2006.02803} {\path{arXiv:2006.02803}},
  \href {https://doi.org/10.1088/1361-6382/abac44}
  {\path{doi:10.1088/1361-6382/abac44}}.

\bibitem{Frey:2019fqz}
A.~R. Frey, {Dirac branes for Dirichlet branes: Supergravity actions}, Phys.
  Rev. D 102~(4) (2020) 046017.
\newblock \href {http://arxiv.org/abs/1907.12755} {\path{arXiv:1907.12755}},
  \href {https://doi.org/10.1103/PhysRevD.102.046017}
  {\path{doi:10.1103/PhysRevD.102.046017}}.

\bibitem{Ertem:2012qv}
U.~Ertem, O.~Acik, {Couplings of gravitational currents with Chern-Simons
  gravities}, Phys. Rev. D 87~(4) (2013) 044052.
\newblock \href {http://arxiv.org/abs/1211.3289} {\path{arXiv:1211.3289}},
  \href {https://doi.org/10.1103/PhysRevD.87.044052}
  {\path{doi:10.1103/PhysRevD.87.044052}}.

\bibitem{Witten:1988sy}
E.~Witten, {The search for higher symmetry in String Theory}, Phil. Trans. Roy.
  Soc. Lond. A 329 (1989) 349--357.

\bibitem{MOTZ-2004}
P.~Mora, R.~Olea, R.~Troncoso, J.~Zanelli, {Finite action principle for
  Chern-Simons AdS gravity}, Journal of High Energy Physics 06 (2004) 036.
\newblock \href {http://arxiv.org/abs/hep-th/0405267}
  {\path{arXiv:hep-th/0405267}}, \href
  {https://doi.org/10.1088/1126-6708/2004/06/036}
  {\path{doi:10.1088/1126-6708/2004/06/036}}.

\bibitem{MOTZ-2006}
P.~Mora, R.~Olea, R.~Troncoso, J.~Zanelli, {Transgression forms and extensions
  of Chern-Simons gauge theories}, Journal of High Energy Physics 02 (2006)
  067.
\newblock \href {http://arxiv.org/abs/hep-th/0601081}
  {\path{arXiv:hep-th/0601081}}, \href
  {https://doi.org/10.1088/1126-6708/2006/02/067}
  {\path{doi:10.1088/1126-6708/2006/02/067}}.

\bibitem{Mora20141}
P.~Mora, {Action Principles for Transgression and Chern-Simons AdS Gravities},
  Journal of High Energy Physics 11 (2014) 128.
\newblock \href {http://arxiv.org/abs/1407.6032} {\path{arXiv:1407.6032}},
  \href {https://doi.org/10.1007/JHEP11(2014)128}
  {\path{doi:10.1007/JHEP11(2014)128}}.

\bibitem{Izaurieta:2011fr}
F.~Izaurieta, E.~Rodriguez, {On eleven-dimensional Supergravity and
  Chern-Simons theory}, Nucl. Phys. B 855 (2012) 308--319.
\newblock \href {http://arxiv.org/abs/1103.2182} {\path{arXiv:1103.2182}},
  \href {https://doi.org/10.1016/j.nuclphysb.2011.10.012}
  {\path{doi:10.1016/j.nuclphysb.2011.10.012}}.

\bibitem{Schutz}
B.~F. Schutz, Geometrical Methods of Mathematical Physics, Cambridge University
  Press, 1980.
\newblock \href {https://doi.org/10.1017/CBO9781139171540}
  {\path{doi:10.1017/CBO9781139171540}}.

\bibitem{Witten-bosonization}
E.~Witten, {Nonabelian Bosonization in Two-Dimensions}, Commun. Math. Phys. 92
  (1984) 455--472.
\newblock \href {https://doi.org/10.1007/BF01215276}
  {\path{doi:10.1007/BF01215276}}.

\bibitem{Deser:1981wh}
S.~Deser, R.~Jackiw, S.~Templeton, {Topologically Massive Gauge Theories},
  Annals Phys. 140 (1982) 372--411, [Erratum: Annals Phys. 185, 406 (1988)].
\newblock \href {https://doi.org/10.1016/0003-4916(82)90164-6}
  {\path{doi:10.1016/0003-4916(82)90164-6}}.

\bibitem{Witten:1983tw}
E.~Witten, {Global Aspects of Current Algebra}, Nucl. Phys. B 223 (1983)
  422--432.
\newblock \href {https://doi.org/10.1016/0550-3213(83)90063-9}
  {\path{doi:10.1016/0550-3213(83)90063-9}}.

\bibitem{Zanelli:1994ti}
J.~Zanelli, {Quantization of the gravitational constant in odd dimensional
  gravity}, Phys. Rev. D 51 (1995) 490--492.
\newblock \href {http://arxiv.org/abs/hep-th/9406202}
  {\path{arXiv:hep-th/9406202}}, \href
  {https://doi.org/10.1103/PhysRevD.51.490}
  {\path{doi:10.1103/PhysRevD.51.490}}.

\bibitem{Duff:1993ye}
M.~J. Duff, J.~X. Lu, {Black and super p-branes in diverse dimensions}, Nucl.
  Phys. B 416 (1994) 301--334.
\newblock \href {http://arxiv.org/abs/hep-th/9306052}
  {\path{arXiv:hep-th/9306052}}, \href
  {https://doi.org/10.1016/0550-3213(94)90586-X}
  {\path{doi:10.1016/0550-3213(94)90586-X}}.

\bibitem{Dixon:1992if}
J.~A. Dixon, M.~J. Duff, J.~C. Plefka, {Putting string / five-brane duality to
  the test}, Phys. Rev. Lett. 69 (1992) 3009--3012.
\newblock \href {http://arxiv.org/abs/hep-th/9208055}
  {\path{arXiv:hep-th/9208055}}, \href
  {https://doi.org/10.1103/PhysRevLett.69.3009}
  {\path{doi:10.1103/PhysRevLett.69.3009}}.

\bibitem{Duff:1994vv}
M.~J. Duff, R.~Minasian, {Putting string / string duality to the test}, Nucl.
  Phys. B 436 (1995) 507--528.
\newblock \href {http://arxiv.org/abs/hep-th/9406198}
  {\path{arXiv:hep-th/9406198}}, \href
  {https://doi.org/10.1016/0550-3213(94)00538-P}
  {\path{doi:10.1016/0550-3213(94)00538-P}}.

\end{thebibliography}

\end{document}